\documentclass[]{spie}  %>>> use for US letter paper
\usepackage{amsmath,amsfonts,amssymb}
\usepackage{graphicx}
\usepackage[colorlinks=true, allcolors=blue]{hyperref}
\usepackage[utf8]{inputenc} % To be able to write ⌀ in the first place
\usepackage{fdsymbol}       % Contains a symbol for ⌀ -- use \diameter
\usepackage{newunicodechar} % To write the definition on the next line
\newunicodechar{⌀}{\ensuremath{\diameter}}

\usepackage{newunicodechar}
\newunicodechar{≤}{\ensuremath{\leq}}

\title{Ultraviolet Extinction Sky Survey (UVESS):
A mission concept for probing the interstellar medium in the Milky Way and Local Group galaxies  
}

\author[a]{Joice Mathew}
\author[b,c]{Andrew Battisti}
\author[a]{Israel Vaughn}
\author[d]{Shubhangi Jain}
\author[d]{Rekhesh Mohan}
\author[d]{Jayant Murthy}
\affil[a]{Advanced Instrumentation and Technology Centre, Research School of Astronomy and Astrophysics, Australian National University, Canberra, ACT 2611, Australia}
\affil[b]{Research School of Astronomy and Astrophysics, Australian National University, Canberra, ACT 2611, Australia}
\affil[c]{ARC Centre of Excellence for All Sky Astrophysics in 3 Dimensions (ASTRO 3D), Australia}
\affil[d]{Indian Institute of Astrophysics, Koramangala 2nd Block, Bengaluru, India 560034 }

\authorinfo{Further author information: (Send correspondence to Joice Mathew, Andrew Battisti)\\Joice Mathew: E-mail: joice.mathew@anu.edu.au\\Andrew Battisti: E-mail: andrew.battisti@anu.edu.au}

\begin{document} 
\maketitle

\begin{abstract}
The 2175 Å bump shows considerable variations in its strength, width, and central wavelength when observed along different sightlines in the Milky Way and other galaxies. These variations offer valuable insights into the composition, size distribution, and processing of interstellar dust grains along different sightlines.  This paper introduces a mission concept called UVESS (Ultraviolet Extinction
Sky Survey) aimed at exploring the composition of the interstellar medium (ISM) within both the Milky Way and nearby Local Group Galaxies by mapping the variation of UV extinction curve slopes and the 2175 Å feature across a majority of the sky to gain insights into the makeup of the ISM. Recent advancements in UV instrumentation and technologies pave the way for the development of high-throughput instruments in compact form factors. In this paper, we outline mission science goals and instrument concept tailored for a small satellite-based platform dedicated to the study of UV extinction.
\end{abstract}

\keywords{UV-extincion bump, UV spectroscopic sky survey, ISM, UV and space instrumentation}

\section{Introduction}
\label{sec:intro}  % \label{} allows reference to this section

The 2175 Å extinction bump is a prominent and ubiquitous feature observed in the interstellar extinction curves of the Milky Way and other galaxies. This broad absorption feature, has been a long-standing enigma in astrophysics since its discovery in 1965.\cite{Mathis}. Despite over half a century of intensive exploration, the exact physical origin of the 2175 Å bump remains elusive and has been the subject of ongoing debate. It is widely believed that the bump is associated with carbonaceous materials, such as graphite, non-graphitic carbon, polycyclic aromatic hydrocarbons (PAHs), or other sp\textsuperscript{2}-bonded carbon compounds\cite{Mathis,Ma2020,2020MNRAS.497.2190M}. Graphite grains were among the first proposed carriers due to graphite having a strong absorption feature near 2175 Å arising from sp2-bonded carbon.\cite{Mathis1990, Mathis,Ma2020,2020MNRAS.497.2190M} The strength of the observed bump implies it must be produced by an abundant material, and carbon is one of the most abundant elements in the universe after hydrogen and helium.\cite{Mathis1990} While carbonaceous compounds remain the leading candidate for the 2175 Å bump carrier, %its exact nature remains elusive, and 
further observational and theoretical work is needed to resolve the outstanding issues and definitively identify the carrier.\cite{Mathis}

\begin{figure}[h]
\begin{center}
\includegraphics[scale=0.9]{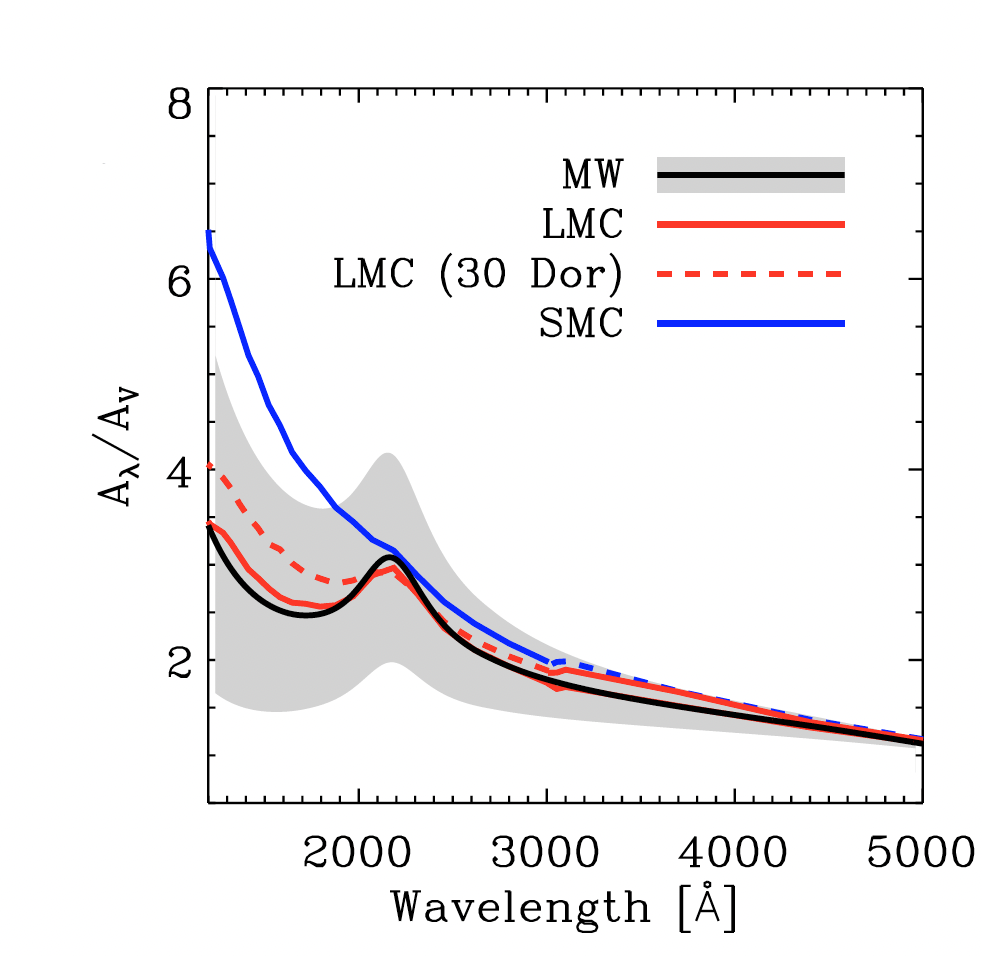}
\end{center}
\caption{Highlighting the large variation of the MW extinction curve (shaded region), taken from the review by Salim and Narayanan (2020). The colored lines show the average extinction curves for the MW and Magellanic Clouds (and 30 Doradus region)\cite{SalimNarayanan2020}}
\label{fig:UV_extinction}
\end{figure}

\begin{figure}[h]
\begin{center}
\includegraphics[scale=0.42]{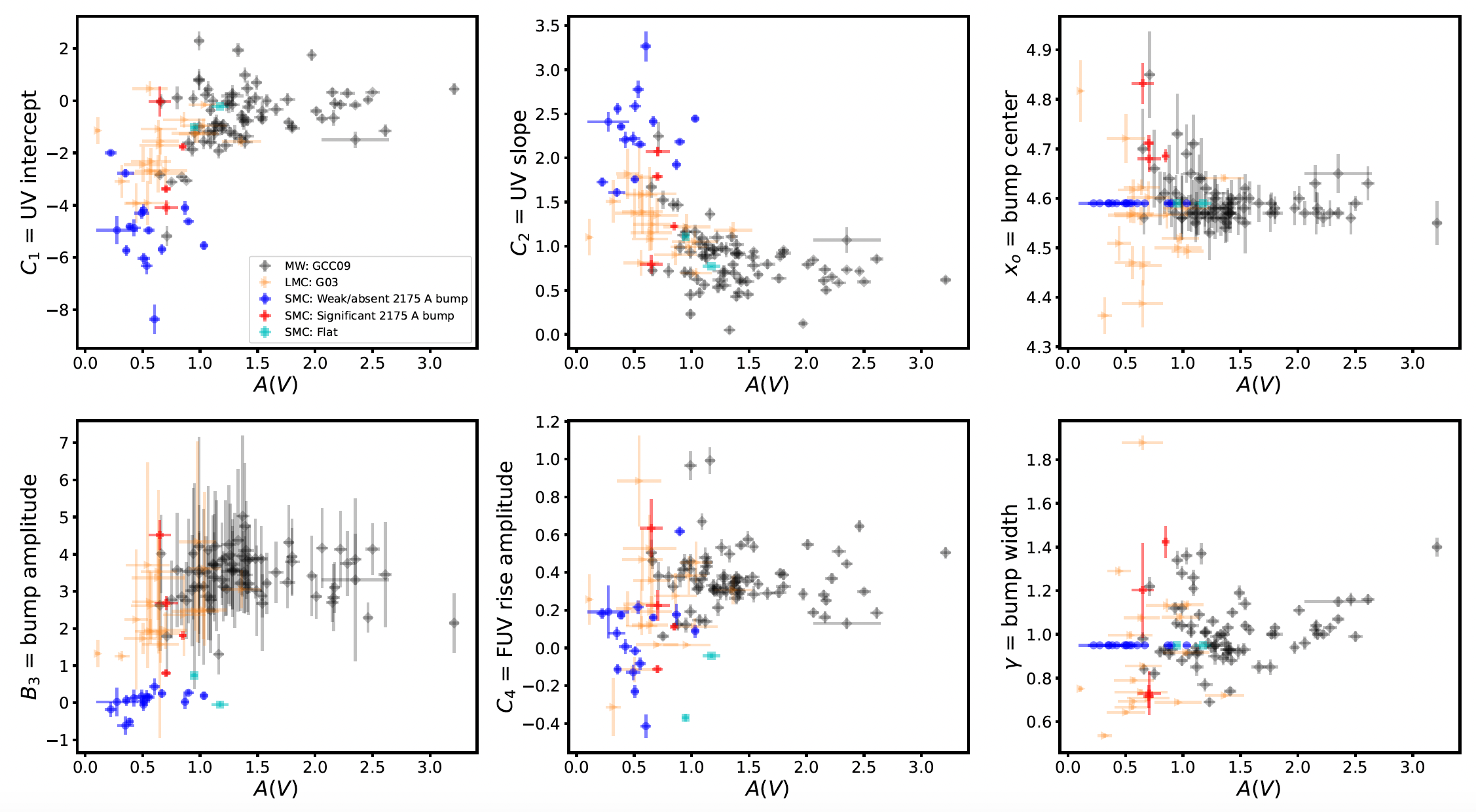}
\end{center}
\caption{Extinction curve properties (parameter definitions follow Fitzpatrick \& Massa 1990) for individual sightlines in the MW and the Magellanic Clouds as a function of dust column density as measured by $A(V)$ ($V$-band extinction) in magnitudes, taken from Gordon et al. (2024)\cite{gordon24}. Past MW studies have primary focused on dustier sightlines that occupy different parameter space than LMC and SMC sightlines, where the latter tend to be less dusty than the MW (lower $A(V)$). With an all-sky UV spectroscopic survey, it will be possible to study MW sightlines extending to low $A(V)$ and bridge this gap. This could determine if these extinction curve parameters follow similar relations despite difference between these systems (e.g., metallicity).}
\label{fig:gordon24_fig10}
\end{figure}

The strength and characteristics of this feature provide valuable insights into the composition, size distribution, and processing of interstellar dust grains along different sightlines. Variations in the bump's strength, width, and polarization properties can be used to trace the evolution of dust grains and their interactions with the local environment. The 2175 Å bump exhibits significant variations in its strength, width, and central wavelength when observed along different sightlines in the Milky Way and other galaxies.\cite{Mathis,2020MNRAS.497.2190M} These variations are thought to be related to differences in the composition, size distribution, and processing of the dust grains responsible for the feature, as well as the local environmental conditions such as radiation fields, supernova rates, and dust production and destruction mechanisms.\cite{Mathis,Ma2020}. The variation of the Milky Way extinction curve, as well as the average curves for the Magellanic Clouds and 30 Doradus region, are shown in Figure~\ref{fig:UV_extinction}. The variation in extinction curve properties along individual lines of sight for these systems are shown in Figure~\ref{fig:gordon24_fig10}.
Observations have also revealed a small but detectable polarization excess associated with the 2175~Å bump, suggesting that the carrier grains are non-spherical and partially aligned with the interstellar magnetic field. This finding provides valuable constraints on the nature and properties of the carrier grains.

The study of the 2175~Å extinction bump holds significant importance in astrophysics as it serves as a powerful diagnostic tool for probing the composition, evolution, and physical properties of interstellar dust grains. It also plays a crucial role in tracing the dust-to-gas ratio, metallicity, and magnetic field structure in various astrophysical environments, from the Milky Way to distant galaxies across cosmic time.\cite{Mathis,Ma2020,Battisti2020}
Additionally, a detailed study of the UV extinction bump reveals how dust interacts with the interstellar environment, shedding light on processes like grain growth, destruction, and the formation of complex organic molecules. This contributes to a comprehensive understanding of the ISM's lifecycle and the interplay between dust and gas.
%. Despite extensive research efforts, the 2175~Å extinction bump continues to be an intriguing and challenging problem in astrophysics, motivating further observational and theoretical investigations to unravel its underlying physical origin and unlock the secrets of interstellar dust grains.
%%
%The UV extinction bump at 2175 Å is a significant feature that provides insights into the composition, size distribution, and properties of interstellar dust grains. 

By studying this feature at extragalactic scales, we can better comprehend various astrophysical processes, including star formation, active galactic nuclei (AGN), and galaxy evolution.  
Correcting for the effect of dust attenuation, which is most pronounced at UV wavelengths, is one of the leading sources of uncertainty in our ability to derive many galaxy properties from photometric datasets (e.g., stellar masses, star formation rates).\cite{SalimNarayanan2020} 
Accurate corrections are vital for obtaining true luminosities and colors of celestial objects, thereby ensuring precise distance measurements and understanding the intrinsic properties of stars and galaxies. 
Dust attenuation and the 2175\AA\ feature can affect the accuracy of photometric redshifts by up to $\sim$10\%\cite{Battisti2019}, which is higher than the requirements for dark energy surveys being done with the \textit{Euclid} and \textit{Roman} space telescopes that require photometric redshift accuracy to $<$5\% to trace large-scale structure.\cite{Euclid2024}

%Moreover, UV extinction studies are relevant to various areas, including the study of active galactic nuclei (AGN), starburst galaxies, and early star formation stages. 

A dedicated small-scale UV spectroscopic mission to study the UV extinction bump is crucial for advancing our understanding of the interstellar medium (ISM). 
At ANU, in partnership with international collaborators, we are developing UVESS (Ultra Violet Extinction Sky Survey), a small satellite UV spectroscopic mission. UVESS will map the variability in extinction curves by acquiring near-UV (140-270 nm) spectra from a large number of target stars.
Such a mission allows for precise measurements of extinction curves, which are essential for improving models used to correct astronomical observations for dust extinction.
Developing dedicated small-scale mission fosters technological advancements in UV instrumentation and spectroscopy. It also provides opportunities for training and involving the next generation of astrophysicists and engineers in cutting-edge research, thereby contributing to the broader scientific and technological community. 

\section{Science Cases}
The 2175 Å extinction bump has significant implications for both Galactic and extragalactic astronomy, as it serves as a powerful diagnostic tool for studying the properties of interstellar dust and the interstellar medium (ISM) in various astrophysical environments, and can moderately affect photometric dust attenuation corrections (e.g., $\sim$50\%) depending on the rest-frame wavelength and width of the filter.

\subsection{Interstellar Medium}

The strength and characteristics of this feature provide valuable insights into the composition, size distribution, and processing of interstellar dust grains along different sightlines. Variations in the bump's strength, width, and polarization properties can be used to trace the evolution of dust grains and their interactions with the local environment. The strength of the 2175 Å bump relative to the total extinction ($A_{bump}/A_V$) is found to correlate with the dust-to-gas ratio and metallicity in different environments.\cite{Mathis}. This relationship allows the 2175 Å bump to be used as a diagnostic tool for studying the dust content and chemical enrichment in various astrophysical systems, from the Milky Way to external galaxies and even high-redshift objects. The observed properties of the 2175 Å bump, such as its central wavelength, width, and polarization behavior, provide stringent constraints for theoretical models of interstellar dust grains.\cite{Andersson2022}

%The 2175 Å extinction bump serves as a powerful diagnostic tool for studying the composition, evolution, and physical properties of interstellar dust grains, as well as tracing the dust-to-gas ratio, metallicity, and magnetic field structure in various astrophysical environments. 
The ubiquitous presence and unique characteristics of the 2175 Å extinction bump make it a valuable probe for advancing our understanding of the interstellar medium and the role of dust in astrophysical processes.\cite{Andersson2022}
By comparing a large-area UV spectroscopic survey, where the 2175A feature can be studied in extinction, to large-area mid-IR surveys (e.g., GLIMPSE survey\cite{Churchwell2009} with \textit{Spitzer} or SPHEREx\cite{Crill2020} in the near-future), where PAH abundances can be inferred from emission, it will be possible to examine the correlation of these dust grains in a statistical manner and determine the likelihood that they arise from the same grain.

\subsection{Implications on extragalactic Astronomy}

In external galaxies, the presence and strength of the 2175 Å bump in the attenuation curves provide \textit{indirect} insights into the dust properties and their dependence on factors like star formation activity, inclination, and dust geometry.\cite{SalimNarayanan2020} 
The reason this is indirect is because it is very difficult to link attenuation curves (i.e., effects of dust on unresolved stellar populations) to physical properties of dust due to degeneracies with geometric and optical depth effects on the behavior of the spectral energy distribution.\cite{SalimNarayanan2020}
Thus, assuming reasonable assumptions regarding star/dust geometry can be made, variations in the bump's strength, width, and polarization properties can be used to indirectly trace the evolution of dust grains and their interactions with the local environment, such as radiation fields, supernova rates, and the production and destruction mechanisms of the carrier grains.\cite{Kashino_2021, Shivaei2022} 
%The strength of the 2175 Å bump relative to the total extinction is found to correlate with the dust-to-gas ratio and metallicity in different regions of the Milky Way.\cite{Shivaei2022} This relationship allows the bump to be used as a diagnostic tool for studying the dust content and chemical enrichment in various Galactic environments, from diffuse clouds to dense molecular clouds. 
The bump strength in star-forming galaxies at high redshifts is found to correlate with stellar mass, specific star formation rate, and polycyclic aromatic hydrocarbon (PAH) emission, suggesting a connection between the bump carrier grains and the recent star formation history and dust processing mechanisms.\cite{Kashino_2021}. By studying the 2175 Å bump in galaxies across cosmic time and different environments, we can gain insights into the formation, evolution, and destruction processes of interstellar dust grains, as well as the role of dust in galaxy evolution and the enrichment of the interstellar medium with heavy elements.\cite{2020MNRAS.497.2190M}

\section{Instrument Overview}

\begin{figure}[h]
\begin{center}
\includegraphics[scale=0.8]{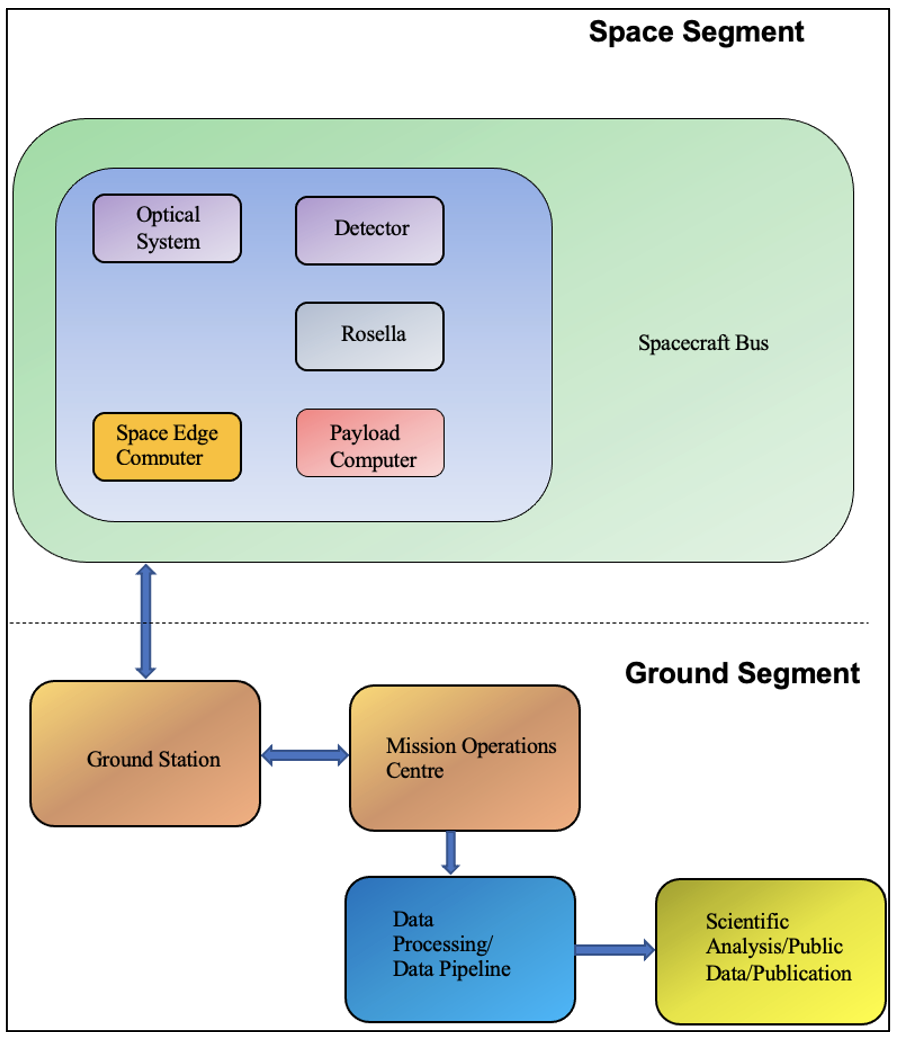}
\end{center}
\caption{Mission architecture}
\label{fig:Mission architecture}
\end{figure}

The proposed instrument is an ultraviolet (140-270 nm) spectrograph designed to fit into small-sat form factor volume ($\sim$ 300 x 200 x 200 mm). Fig ~\ref{fig:Mission architecture} illustrates the mission architecture, highlighting various subsystems within both the ground and space segments. The instrument features a compact UV spectrograph paired with an MCP-based detector system. The phosphor screen of the MCP will be read out using a fast frame CMOS sensor with subsequent data handling by the Rosella detector control system. The data will be further processed using the space edge computer and will be transferred to the spacecraft bus for storage and transmission using the payload computer. The primary instrument specifications are shown in Table ~\ref{table:UVESS instrument details}. 

\begin{table}[h]
\caption{Instrument details}
\begin{center}
\begin{tabular}{ll}
\hline 

\rule[-1ex]{0pt}{3.5ex} Telescope type & Modified RC \\
\rule[-1ex]{0pt}{3.5ex} Band of operation & $140-270$ nm\\
\rule[-1ex]{0pt}{3.5ex} Spectral resolution & $1$ nm\\
% \rule[-1ex]{0pt}{3.5ex} Spectrograph & Czerny Turner \\

% \rule[-1ex]{0pt}{3.5ex} Weight & $ 2 $ kg \\
\rule[-1ex]{0pt}{3.5ex} F number & 4.5 \\
\rule[-1ex]{0pt}{3.5ex} Aperture   &  150$\times$86 mm  \\
\rule[-1ex]{0pt}{3.5ex} Focal length & 800 mm\\
\rule[-1ex]{0pt}{3.5ex} Field of view & $\sim$ One degree \\
\rule[-1ex]{0pt}{3.5ex} Grating & Toroidal (1200 lines/mm) \\
\rule[-1ex]{0pt}{3.5ex} Detector & Solarblind MCP (40 mm diameter) \\
\rule[-1ex]{0pt}{3.5ex} Dimension (L$\times$W $\times$H)& $200\times 100\times100$ mm \\
% \rule[-1ex]{0pt}{3.5ex} Pixel scale & $1.2^{\prime\prime}$/pixel \\

% \rule[-1ex]{0pt}{3.5ex} Limiting magnitude & 12 AB\\
% \rule[-1ex]{0pt}{3.5ex} Bright limit & 2 AB\\
% \rule[-1ex]{0pt}{3.5ex} Minimum exposure time & 0.08 sec \\
\hline
\end{tabular}
\label{table:UVESS instrument details}
\end{center}
\end{table}

\begin{figure}[h]
\begin{center}
\includegraphics[scale=0.30]{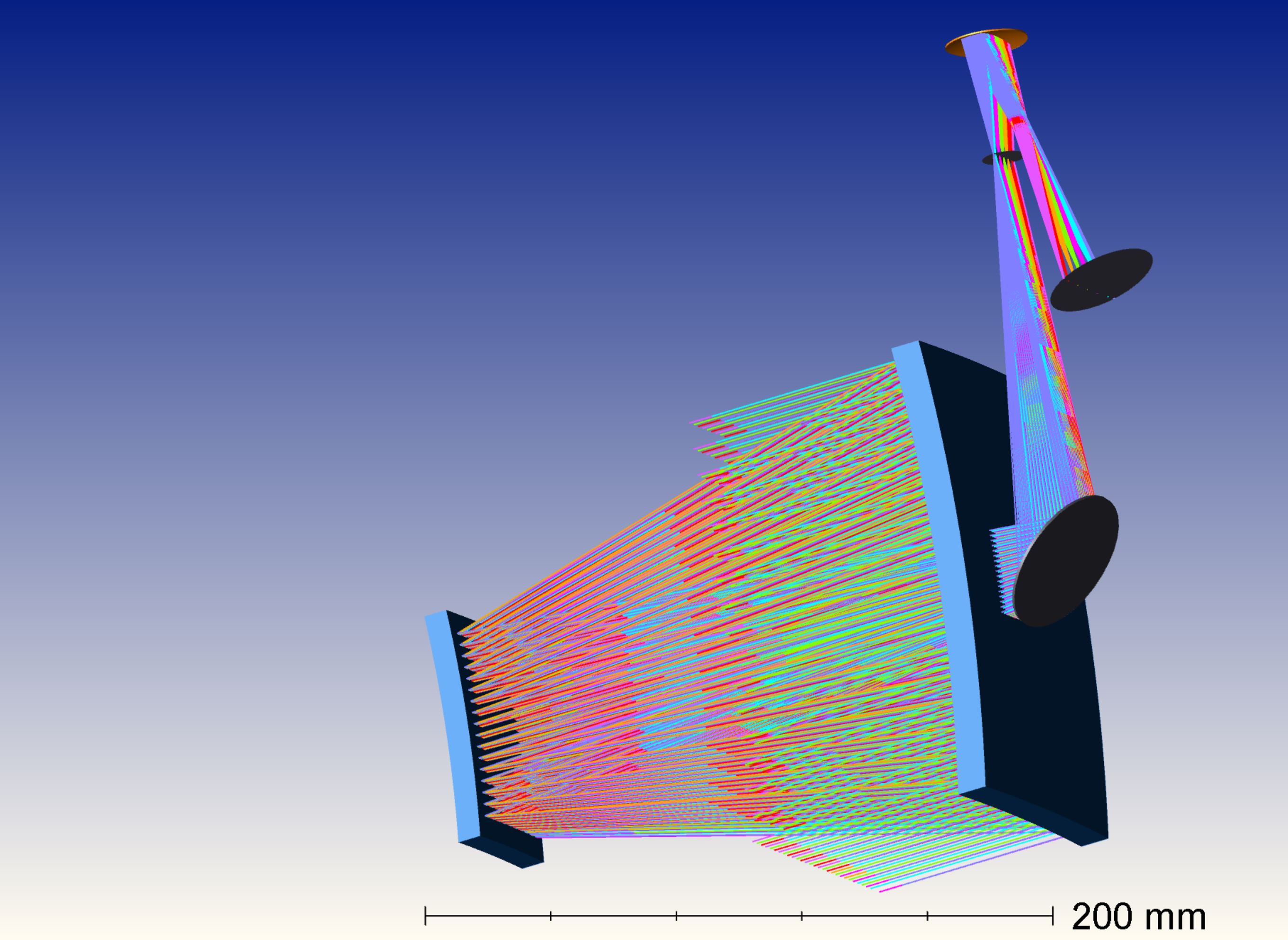}
\end{center}
\caption{Instrument preliminary optical layout}
\label{fig:Instrument preliminary optical layout}
\end{figure}

\begin{figure}[h]
\begin{center}
\includegraphics[scale=0.35]{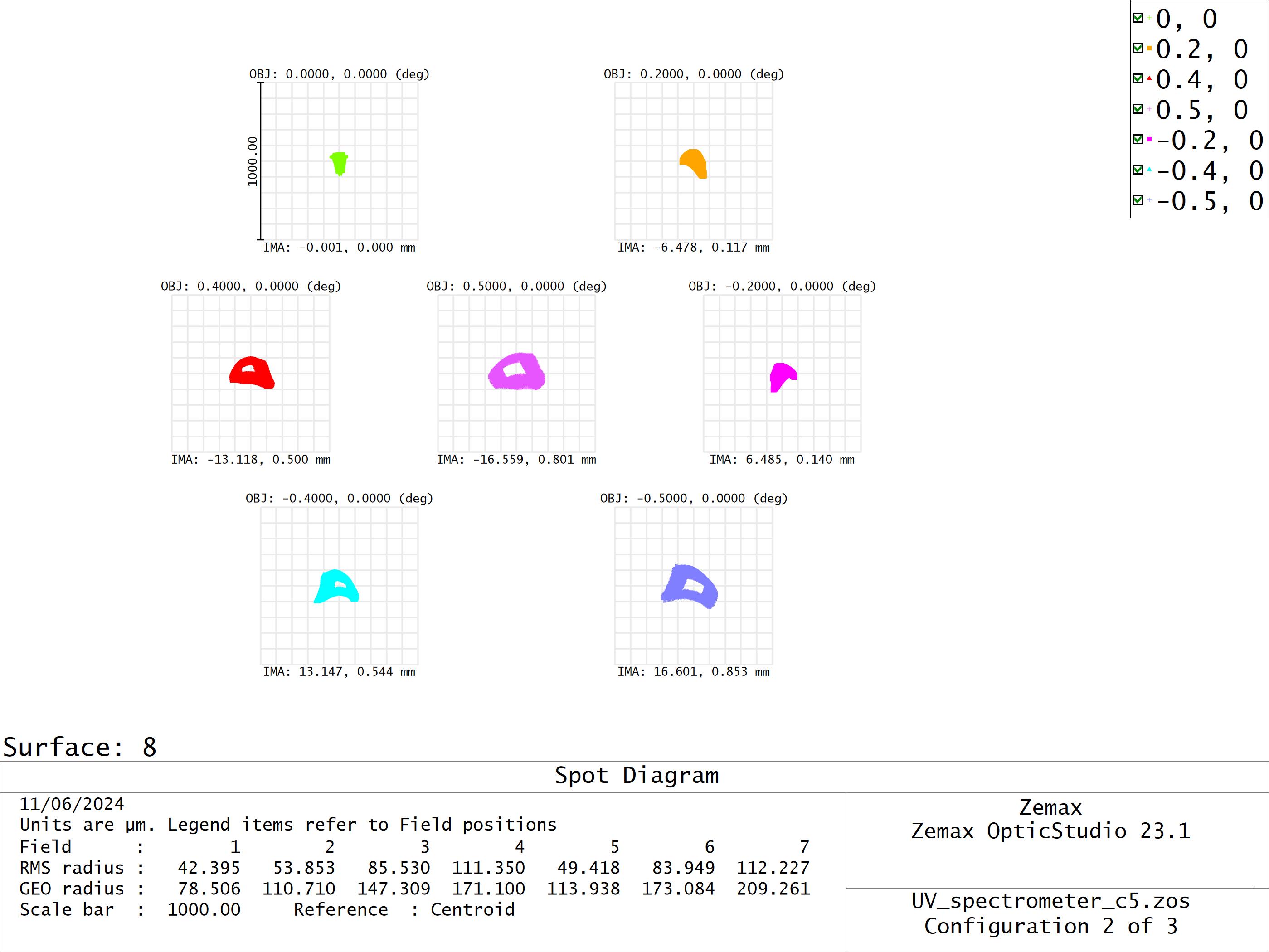}
\end{center}
\caption{Spot diagram for different fields}
\label{fig:Spot diagram for different fields}
\end{figure}

\begin{figure}[h]
\begin{center}
\includegraphics[scale=0.35]{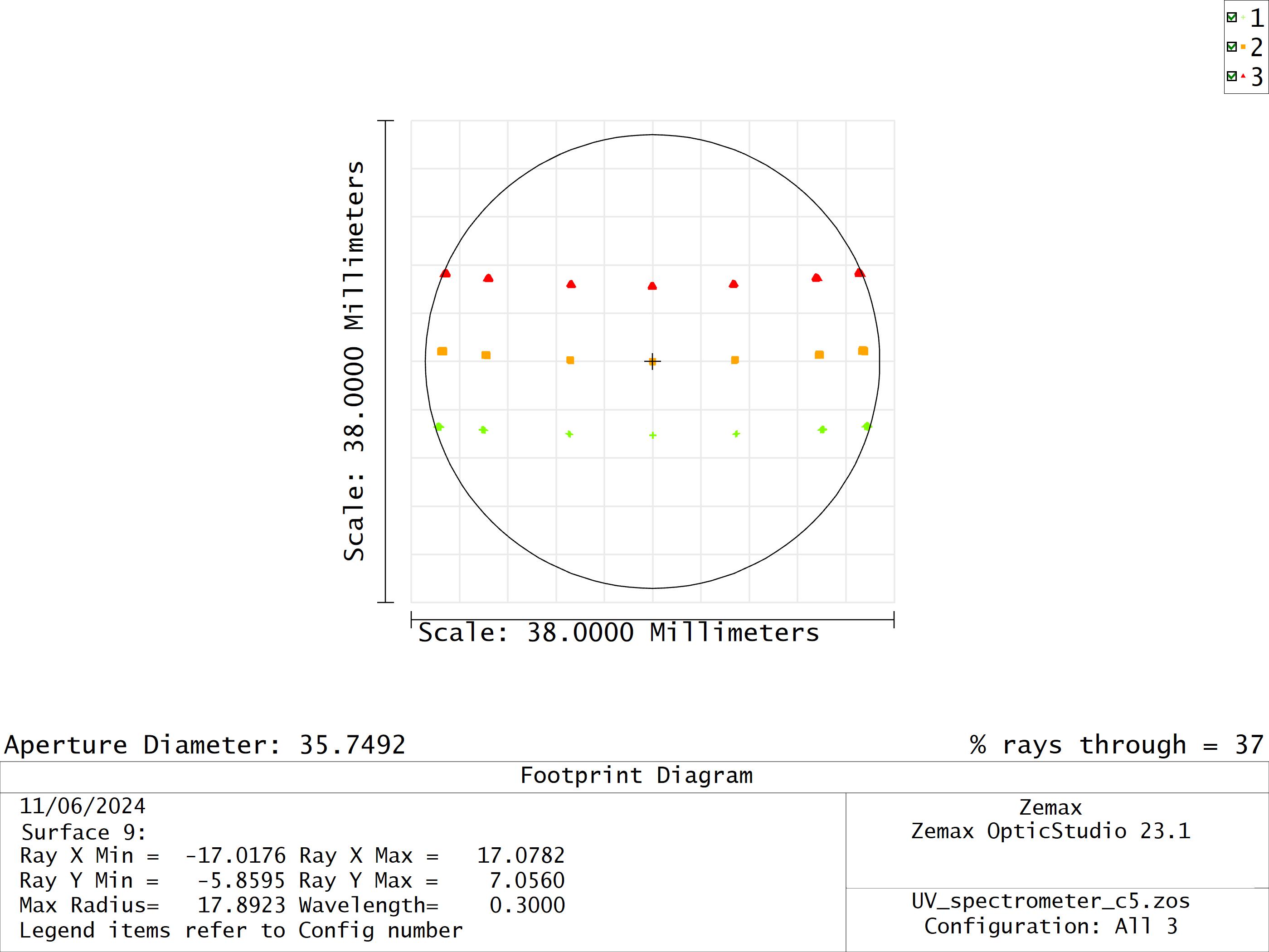}
\end{center}
\caption{Foot print diagram}
\label{fig:Foot print diagram}
\end{figure}

\subsection{Optics}
The UVESS system is designed to maximize the telescope collecting area and utilise high-quality optical elements, including UV-grade mirrors and toroidal grating. The UVESS aperture is rectangular in shape, featuring a 15.0 × 8.6 cm. UVESS's compact UV spectrograph features an 800 mm focal length with 1200 lines/mm toroidal grating, covering 140-270 nm and achieving $R\sim2500$. It has a 9556 mm² collecting area, an f/4.5 aperture, and can be further compacted to fit within 4U. The preliminary optical layout is shown in Figure~\ref{fig:Instrument preliminary optical layout}. The spot diagrams and footprint for the optical system is shown in Figure~\ref{fig:Spot diagram for different fields} and Figure~\ref{fig:Foot print diagram} respectively.  The reflective surfaces are set to be coated with a reflective aluminium (Al) coating and a protective layer of magnesium fluoride (MgF2), ensuring reflectivity exceeding 85\% within the operational near-ultraviolet (NUV) band. This MgF2 layer acts as a barrier against aluminium oxidation, crucial because even thin oxide layers, measuring just a few nanometres, can compromise the reflectivity of the Al coating in UV environments.

\begin{figure}[h]
\begin{center}
\includegraphics[scale=0.95]{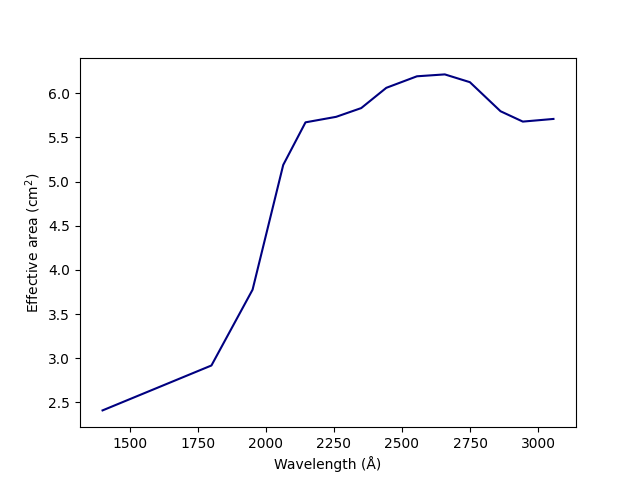}
\end{center}
\caption{UVESS Effective area}
\label{fig:UVESS Effective area}
\end{figure}

UVESS effective area is give by-
\begin{equation}
A(\lambda )=A_{T} M(\lambda )^3 G(\lambda ) \eta (\lambda )
\end{equation}

where $A_{T}$, $M(\lambda )$, $G(\lambda )$ $\eta(\lambda )$,  are the effective geometrical collecting area of the telescope, the reflectivity of mirrors, efficiency of grating and quantum efficiency of the MCP detector respectively. Figure~\ref{fig:UVESS Effective area} shows the calculated effective area of UVESS as a function of wavelength.

\subsection{Detector}

The instrument will use MCP340 assembly from Photek as the detector. It is a 40-mm Z-stack Micro Channel Plate (MCP) with a solar blind Caesium-Telluride photocathode. It is sensitive to UV radiation from 140 to 270 nm and achieves a maximum quantum efficiency (QE) of approximately 20\% within its operating band. The MCP operates using a high voltage power supply (model FP6329) that provides the necessary voltages of 5500 V, 2400 V, and -200 V. The phosphor screen behind the MCP will be imaged using a CMOS sensor and this sensor will be readout using the Rosella detector control system. The LUPA1300-2 CMOS is a 1.6 MP detector array from ON Semiconductor that meets our requirements for faster readout and a large sensor size. It features 14 µm pixels with an 18 × 14 mm size, matching the output diameter of the taper and resulting in fewer aberrations when interfaced with the MCP. The sensor supports high-speed clocks of up to 315 MHz, enabling a maximum frame transfer rate of up to 500 fps. Additionally, it supports windowed and sub-sampled readouts, allowing for a temporal resolution of 2 milliseconds at full resolution (1.3 MP) and 10 µs for the smallest window. The Rosella system, with its maximum readout rate of 1 KHz, is well-suited for this detector readout. The acquired images will be processed in real-time by the space edge computer to detect events, perform centroiding, and conduct subsequent data processing.

\begin{figure}[h]
\begin{center}
\includegraphics[scale=0.60]{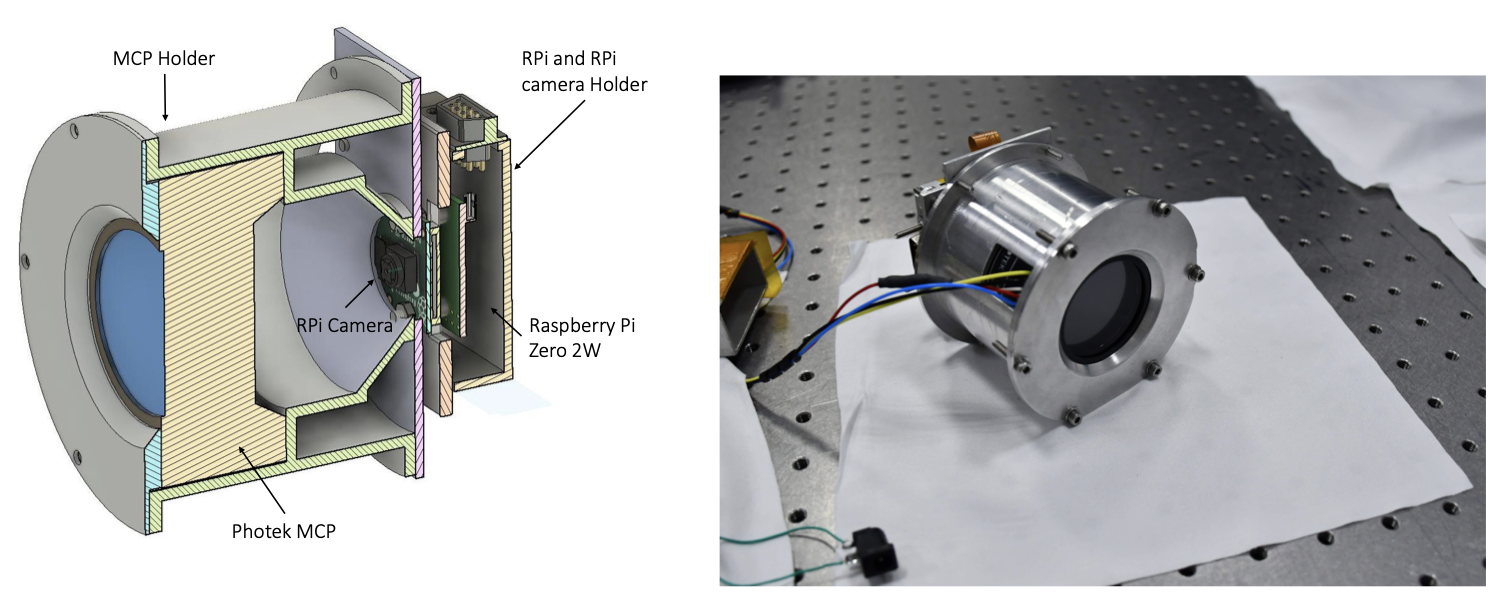}
\end{center}
\caption{MCP based UV photon counting detector\cite{iia_detector}}
\label{fig: MCP based UV photon counting detector}
\end{figure}

\subsection{Rosella}
\begin{figure}[h]
\begin{center}
\includegraphics[scale=0.6]{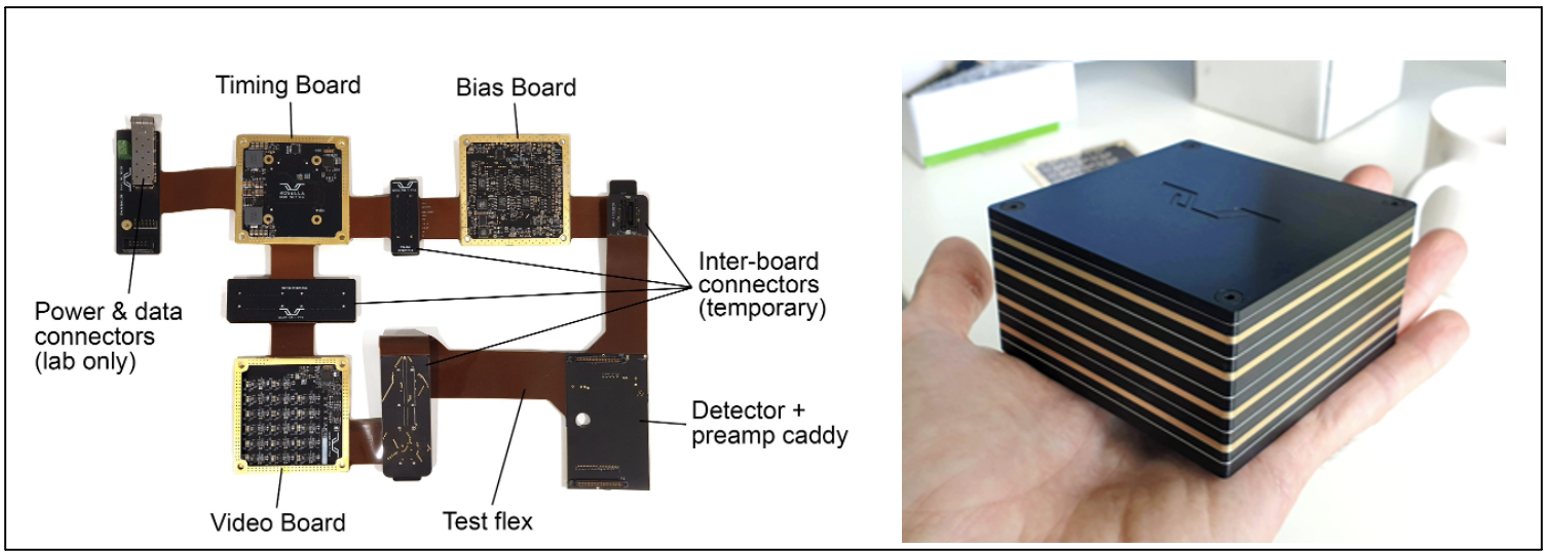}
\end{center}
\caption{Left: Rosella engineering model ‘FlatSat’. Right: Rosella 0.5 U enclosure mock-up\cite{Rosella}}
\label{fig:Rosella}
\end{figure}

The CMOS focal plane assembly will be electronically integrated with the Rosella control electronics currently at technology readiness level (TRL) 4-5 \cite{Rosella}.  Rosella FEE is a modular and compact detector controller for space applications under development by ANU in Canberra.  This high-performance FPGA-based readout system can be configured to interface with a wide range of visible and infrared CMOS detectors.  The system is highly configurable to deliver high-performance frame rate, low noise level and bespoke windowed readout.  Rosella flatsat is shown in Figure~\ref{fig:Rosella}.
The Rosella timing board is responsible for managing the entire system, including clock pattern generation, bias configurations, analogue-to-digital converter triggering, image processing, and communication with an external payload computer via a standard protocol.  Rosella provides a simple and low-level interface to the Instrument Control Unit (ICU) for high-level tasking, as well as direct output to GPU systems to support onboard AI analysis and real-time value-added data analysis.

\subsection{Payload Computer Unit (PCU) }
We will be using a flight-proven Volkh processor from Infinity Avionics as the payload computer unit (PCU) for this mission.  The PCU provides the interface between the satellite platform management module and the payload. Volkh processor is a reconfigurable processing platform for space applications that can be used as a spacecraft onboard computer as well as a payload processor. Unique FPGA-based architecture enables reconfiguring communication interfaces and GPIOs based on system requirements. FPGA SoC-based processor comes in a small form factor making it a flexible option for a platform/payload processor. Its small form factor, lightweight and data storage options including SEU tolerant MRAM and low-power DDR memory make it a reliable choice for the proposed payload.

\section{Challenges in Studying the 2175 Å Bump}

Observing the 2175 Å feature from ground-based observatories is virtually impossible due to the Earth's atmosphere, which absorbs most of the UV radiation. This necessitates the use of space-based observatories or high-altitude balloon missions. However, both options come with significant logistical and financial challenges. Space missions, while offering the best observational conditions, are expensive and complex to design and launch. Balloon missions, though more cost-effective, provide limited observation time and are subject to atmospheric disturbances even at high altitudes. One of the primary challenges in studying the 2175 Å bump is achieving high sensitivity in the UV range. UV detectors and optical elements generally have lower efficiencies compared to their visible and infrared counterparts. The need for high-resolution and high-sensitivity instruments necessitates ongoing advancements in detector technology, such as improved charge-coupled devices (CCDs) and microchannel plates (MCPs), as well as enhanced UV coatings that maximize reflectivity and transmission.

Precise calibration of UV spectrographs is crucial for accurate measurements. This includes accounting for instrumental effects such as wavelength calibration, flat-fielding, and detector non-linearities. Maintaining the stability of these calibrations over time is also a significant challenge, especially for long-duration missions. Calibration sources (such as CALSPEC data base \cite{Bohlin2014}) and methodologies must be meticulously developed and regularly updated to ensure data accuracy. Extracting the 2175 Å feature from observed spectra involves fitting the extinction curve with accurate models that account for the underlying stellar spectrum and other extinction components. Developing robust models that can account for the variability and complexity of the feature is an ongoing challenge. Moreover, interpreting the results requires a comprehensive understanding of the physical and chemical processes in the interstellar medium (ISM), which are not yet fully understood. With the proposed UVESS mission, we aim to address the challenges associated with exploring the composition of the interstellar medium (ISM) by mapping UV extinction curve slopes and the 2175 Å feature across vast regions of the sky.

% Addressing the challenges in studying the 2175 Å bump will require a combination of technological innovation, strategic planning, and interdisciplinary collaboration. Continued advancements in UV detector technology and optical coatings are essential for improving the sensitivity and resolution of UV spectrographs. Innovations in lightweight, high-throughput instruments will enhance the capability of both space-based and balloon-borne observatories. Future directions in this field are poised to significantly enhance our understanding of interstellar dust and its role in the cosmos, offering new insights into the composition and behavior of the ISM across different galactic environments. 

\section{Polarisation}
Observations of polarization in the UV wavelength range have revealed a small polarization excess associated with the prominent 2175 Å extinction bump.\cite{Andersson2022} This finding has important implications as it provides valuable insights into the nature and properties of the carrier grains responsible for this prominent interstellar feature. The first evidence of polarization excess came from UV spectropolarimetric observations of the heavily reddened stars HD 197770 and HD 147933-4 using the Hubble Space Telescope's Faint Object Spectrograph (HST/FOS).\cite{Somerville1994} These observations showed a distinct polarization excess indicating that the grains responsible for the bump feature exhibit some degree of non-spherical shape and alignment with the interstellar magnetic field.

The small polarization excess associated with the 2175 Å bump has important implications for the nature of the carrier grains. It suggests that the carrier grains are likely to be small (size $\leq$ 0.01 $\mu$m) and have an irregular or non-spherical shape \cite{Mathis}. The poor polarization efficiency could be due to the grains being only partially aligned or having a complex shape that reduces the overall alignment efficiency. Theoretical models have proposed that the 2175 Å bump carrier could be carbonaceous grains containing sp\textsuperscript{2}-bonded structures, such as graphite or polycyclic aromatic hydrocarbons (PAHs) \cite{Somerville1994, Lin2023}. The observed polarization excess supports the idea that the carrier grains are non-spherical carbonaceous particles, although their exact composition and structure remain uncertain. We will perform a feasibility analysis to assess the integration of an additional polarization measurement capability, along with spectroscopy, into the current instrument for the purpose of studying the extinction bump.

\section{Observation Strategy}
The mission will conduct UV spectroscopic observations of stars within the Milky Way galaxy, combining the observed stellar spectra with data from astrometric surveys (Gaia) and synthetic photosphere models to construct a three-dimensional map of extinction sightlines to study extinction and dust grain properties. The stellar type and metallicity information will refine the target selection, and the resulting 3D maps will be correlated with infrared and radio maps to provide a multi-wavelength view of the carrier's relationship with various stellar structures and ISM environments. For galaxies in the Local Group, larger galactic regions will be observed to map the carrier distribution on galaxy-wide scales in 2D, utilizing simple stellar population frameworks to model the expected UV emission in the absence of extinction\cite{Calzetti2021}. An iterative, hierarchical observation strategy will be adopted, targeting O-B3 stars providing bright UV emission, B3-A stars with lower UV emission but more evenly distributed, and bright star-forming regions in Local Group galaxies exhibiting diffuse UV emission.

\section{Summary}
The UVESS mission focuses on studying the UV extinction bump to enhance our understanding of the ISM and celestial object formation. Utilizing a compact UV spectrograph, it aims to make precise measurements of this phenomenon across different stars. By combining UV spectroscopic observations with data from astrometric surveys and synthetic models, UVESS will construct three-dimensional maps of extinction sightlines, refining our understanding of dust properties and their relationship with various stellar structures and ISM environments. Additionally, the mission extends its scope to galaxies within the Local Group, employing observational strategies targeting different star types and regions.

\section{Acknowledgment}

The authors humbly acknowledge and pay their respects to the traditional custodians of the land upon which the majority of this project has taken place: the Ngunnawal and Ngambri peoples. In doing so, we honour all First Nations people, recognizing their enduring resilience, rich culture, and invaluable contributions.
% References
\bibliography{report} % bibliography data in report.bib
\bibliographystyle{spiebib} % makes bibtex use spiebib.bst

\end{document}